\documentclass[aps,prl,twocolumn,showpacs]{revtex4} 
\usepackage{graphicx}
\usepackage{amsmath}
\usepackage{amssymb}

\font\FermiSmallfont=cmssq8 scaled 1200
\def\LANLppthead#1#2{
\null 
\begin{center}\vskip -1.0truein{\hbox to 7.5truein {
\hfill
\vbox to 1in {\vfill \FermiSmallfont
              \hbox{#1}
              \hbox{#2}
              \vfill}
}}\vskip-0.0truein\end{center}}

\def\photon2{\gamma \gamma}

\def\cv{c_{\rm v}} 
\def\GeV{\ {\rm GeV}} 
\def\cm{{\rm cm}} 
\def\s{{\rm s}}  
  
\def\Aeff{{\cal{A}}_{\rm eff}}

\def\Eth{E_{\rm th}}  
\def\Ngs{N_{\gamma, \rm  s}}

\def\Td{T_d}
\def\sigmav{\langle \sigma  v \rangle}
\def\calD{{\cal D }}
\def\Mm{M_{\rm m}}
\def\MeV{{\rm MeV}}
\def\dmax{ {\cal D}_{\rm max}}
\def\Nm{N_{\rm m}}
\def\zform{z_{\rm form}}

\def\thetamin{\theta_{\rm min}}

\def\Mchi{M_\chi}
\def\Ngth{N_{\gamma, \rm th}}
\def\Msun{M_\odot}
\def\muth{\mu_{\rm th}}

\begin{document}

\LANLppthead {LA-UR 06-3781}{astro-ph/0606208}

\title{Proper motion of $\gamma$-rays from microhalo sources }

\author{Savvas M. Koushiappas} 
\affiliation{T-6, Theoretical Division,
  \& ISR-1, ISR Division, MS B227, Los Alamos National Laboratory, Los
  Alamos, NM 87545, USA; smkoush@lanl.gov }



\begin{abstract}
I discuss the prospects of detecting the smallest dark matter bound structures 
present in the Milky Way by 
searching for the proper motion of $\gamma$-ray sources in the upcoming GLAST 
all sky map. I show that for dark matter particle candidates that couple to photons 
the detection of at least one $\gamma$-ray microhalo source 
with proper motion places a constraint on the couplings and mass of the dark matter 
particle. For SUSY dark matter, proper motion detection implies that the mass of the 
particle is less than 500 GeV and the kinetic decoupling temperature is in the range 
of [4-100] MeV. 
\end{abstract}

\maketitle

In the $\Lambda$CDM cosmological model, $\sim 26 \%$ of the present-day energy density 
of the Universe is in the form of dark matter \cite{Spergeletal} which remains an
outstanding problem in cosmology and particle physics. 
Favored Cold dark matter (CDM) particle candidates (see \cite{BHS05} and 
references therein) are Weakly Interacting Massive Particles (WIMPs). They are 
kept in thermal equilibrium through interactions of the form 
$\chi \chi \leftrightarrow f \bar{f}$, and in kinetic equilibrium through 
$\chi f \leftrightarrow \chi f$.
The temperature at which these particles decouple 
from the radiation fluid is called the freeze-out temperature, while the 
temperature at which they 
decouple from scattering interactions  
is called the kinetic decoupling temperature ($\Td$) 
\cite{Schmid:1998mx,HSS01, CKZ02}.

In the CDM paradigm, 
the first collapsed and virialized dark matter structures, {\it microhalos}, are 
formed at high redshifts\cite{Berezinsky:2003vn,Green:2003un, Green:2005fa}. 
Their minimum mass is set by the RMS dark matter particle 
velocities set at kinetic decoupling; scales smaller than the free-streaming 
scale are smoothed out. 
For the case of supersymmetric (SUSY) cold dark matter (CDM), this mass scale is 
$\Mm \approx 10^{-4} M_\odot \,( \Td / 10 \,\MeV )^{-3}$ \cite{HSS01,Berezinsky:2003vn,LZ05}.
Recent numerical simulations of the formation of microhalos in the context of SUSY CDM 
provided hints on their properties, and on their evolution and 
survival in the hierarchical structure formation scenario \cite{DMS05,DKM06}. Their 
potential detection using $\gamma$-rays has been studied in Refs.~\cite{Pieri:2005pg,DKM06}.
Based on simple analytical studies, 
it has been argued that the abundance of microhalos in the Milky Way can 
be severely reduced due to interactions in early rapid merging processes and 
in stellar and galactic disc interactions \cite{ZSH1,METAL05,Berezinsky:2005py,Green:2006hh}.
These arguments are still under debate and need to be resolved through detailed 
numerical simulation studies. 

Investigating the existence of microhalos in the Milky Way could be very important for two reasons: 
1) A possible detection can provide information about the particle physics properties 
of the dark matter particle, and 2) a measured abundance in the Milky Way halo contains 
information on the hierarchical assembly of dark matter halos at very early times, a 
task that is unattainable by any other method. 
If a fraction of microhalos survive in the present day Milky Way halo, and the 
dark matter particle couples to photons (as is the case for SUSY CDM as well as candidates 
that originate from theories with extra-dimensions),   
they may appear as $\gamma$-ray 
sources in the sky with potentially detectable proper motion \cite{METAL05}. 
A search for proper motion of unidentified $\gamma$-ray sources can provide 
important information on the physics of the dark matter particle and on the 
survival rate of microhalos in the Solar neighborhood. 
In this paper, I present proper motion signals that should be expected in the upcoming 
GLAST $\gamma$-ray experiment (http://glast.gsfc.nasa.gov) for dark matter particle candidates 
that couple to photons. I find that for SUSY CDM particles with mass less than $\Mchi \leq 
500$ GeV and kinetic decoupling temperature of $\Td \sim [4- 100]\MeV$ there will be at least one 
detectable microhalo with proper motion if $\sim 20 \%$ of microhalos at the Solar radius survive 
\cite{Berezinsky:2005py,Green:2006hh}.

{\bf \it Photons from Microhalos:}
The number of photons with energy greater than $\Eth$ 
on Earth from a microhalo at distance $\calD$ is 
\begin{equation}
\label{eq:ns}
\Ngs(E>\Eth)= \frac{1}{4 \pi} 
\frac{N_\gamma \sigmav}{\Mchi^2}
g[\rho(r)] {\cal A}_{\rm eff} \tau_{\rm exp}.
\end{equation} 
Here, 
$\sigmav$ is the annihilation cross section 
of the dark matter particle to photons, $\Mchi$ is the mass of the dark matter particle and 
$N_\gamma = \int_{\Eth}^{\Mchi} [dN_\gamma / dE] dE$ is the total number of photons 
emitted above a threshold energy $\Eth$ and $dN_\gamma / dE$ is the 
spectrum of the emitted photons. 
The effective area of the detector is ${\cal A}_{\rm eff}$, 
and $ \tau_{\rm exp}$ is the 
exposure time. 
The term $ g[\rho(r)]$ contains all the information about the distribution of dark matter in a 
microhalo at a distance $\calD$, 
$ g[\rho(r)] = \calD^{-2}\int \rho^2(r) d^3 r = 2 \pi \int_0^{{\phi_m}} 
\sin \psi  \{ \int \rho^2[r(s,\psi,\calD)] \, ds \} \, d \psi$. 
The integral is performed along the line of sight, for an angular extent of 
$\phi_m =\tan^{-1}[ r_s / \calD]$, where $r_s$ is the scale radius.  
The distribution of mass is $\rho(r)$ and is 
assumed to be described by an NFW profile \cite{NFW96} (for an NFW profile, $\sim$90\% of 
the flux comes within the region inside of $r_s$). 
All microhalos 
are described by the same value of the quantity $ g[\rho(r)]$ if 
all are formed at the same 
redshift, $\zform \approx 70$ \cite{DMS05}. 
Given a concentration parameter $\cv \approx 1$ \cite{DMS05}, a microhalo of mass $\Mm$ is 
described by an NFW profile with a characteristic density 
of $\rho_s \approx 756 \GeV \, \cm^{-3}$ and a scale radius 
$r_s \approx 3 \times 10^{-3} \, {\rm pc}  {\Mm}_{,-6}^{1/3} Z_{70}^{-3}$, 
where ${\Mm}_{,-6}=\Mm / 10^{-6} \, M_\odot$ and $Z_{70}= [(1+\zform)/71]$
For the case of SUSY CDM, 
the number of photons detected on Earth that originate 
within a microhalo region defined by the PSF of GLAST 
can be written as
\begin{equation}
\label{eq:Ns}
\Ngs \approx 50 \left( \frac{ {\Mm}_{,-6}}{\calD^2_{0.05}} \right) 
\left( \frac{ N_\gamma }{30} \right) 
\left( \frac{ \sigmav_{-26}}{{\Mchi}_{,40}^2} \right) \, Z_{70}^3 .
\end{equation}
Here, $\calD_{0.05} = \calD / 0.05 {\rm pc}$, the threshold energy is $\Eth = 3 \GeV$ and  
the Hubble parameter is taken as $h=0.7$. 
In this expression, 
$\sigmav_{-26}= \sigmav / 10^{-26} \cm^3 \s^{-1}$, ${\Mchi}_{,46}= \Mchi / 40 \GeV$, 
and the photon spectrum is taken from Ref.~\cite{BUB97}.
This particular choice of cross section, particle mass and threshold energy 
represents an optimistic approach to the case of SUSY CDM detected with GLAST \cite{KZW04}, 
because the annihilation cross section can be up to a factor of $10^6$ smaller, while the 
mass can be as high as numerous TeV. 
However, for dark matter candidates that arise from stable 
Kaluza-Klein excitations in models with universal extra dimensions  
the annihilation cross section to photons can be much higher \cite{BHS05}.

In  order  to  detect  the  $\gamma$-ray signal from 
microhalos, the source signal must be larger  than the 
threshold number of photons for detection, i.e., $\Ngs \ge \Ngth$, where 
$\Ngth \approx 6 $ for a GLAST orbit-averaged effective area of $\Aeff \approx 2 \times 10^3 \cm^2$ 
\cite{Ritz} and an exposure time of 2 years. 
Given a microhalo of mass $\Mm$, the maximum distance from the Sun $\dmax$ at which a 
microhalo can be detected is obtained by setting Eq.~\ref{eq:ns} equal to $\Ngth$. I define the 
``visibility volume'' of a microhalo of mass $\Mm$ to be the volume defined through $\dmax$. 

Microhalos formed at $\zform \approx 70$ 
will merge to form present day dark matter halos. A fraction  
of the Milky Way halo mass perhaps remains in the form of microhalos 
formed at very high redshifts. Simulations show that 
the subhalo mass function in Milky Way 
size halos is $ dN/dM \sim M^{-2}$, with about 10\% of the 
mass of the Milky Way in halos of mass greater than $10^7 \Msun$. 
Analytical arguments based on approximations for the survival probability 
of microhalos in an evolving parent halo \cite{Berezinsky:2005py}, 
as well as preliminary numerical 
work on the mass function of subhalos down to microhalo scales \cite{DKM06}, find
the form of the subhalo mass function is preserved down to microhalo scales. 
Thus, the amount of mass per logarithmic mass interval 
is a constant, and depends only on the survival probability of microhalos.  
If approximately 20\% of microhalos survive \cite{Berezinsky:2005py} in the 
Solar neighborhood (with perhaps a small mass dependence on microhalo 
mass \cite{Green:2006hh}), then the fraction of the local mass in microhalos 
is roughly of order 0.2\%. 
I define the quantity $\xi$ to be the fraction of the dark matter density in 
the solar neighborhood that is in the form of microhalos, i.e., $\xi=0.002$. 
If there are $\Nm$ microhalos in a logarithmic interval in $\Mm$, 
the number of microhalos 
per radial distance interval from the Sun is $d\Nm/d \calD = 4 \pi \xi \calD ^2 \rho(r_\odot) / \Mm$, 
and the total number of microhalos per logarithmic mass interval in $\Mm$ 
within a visibility volume is 
$N_{\rm m} = \int_0^{{\dmax}} [d\Nm/d \calD ] d \calD$.  
The Sun is located at $r_\odot \approx 8$ kpc from the 
galactic center, and the local dark matter density is 
$\rho(r_\odot)= 0.01 \, M_\odot \, {\rm pc}^{-3}$ \cite{HF00}.


{\it Microhalo proper motion}: 
I assume microhalos obey a Maxwell-Boltzmann velocity distribution function set by the 
virial temperature of the Milky Way halo. On the Earth's frame, the distribution function 
can be written as $f(v) dv = [v \zeta(v) / \pi ^{1/2} v_e v_0 ] dv$. 
where, $\zeta(v) = {\rm exp}[(v_1/v_0)^2] - {\rm exp}[(v_2/v_0)^2]$, $v_1=v-v_e$ and $v_2=v+v_e$. 
Here, $v_0\approx 220$ km s$^{-1}$ is the velocity of the Sun in the 
frame of the Galaxy, and $v_e=1.12 v_0$ is the speed of the Earth 
(the very small annual modulation of the Earth's velocity is neglected).
A microhalo at a distance $\calD$ from the Sun, and with a velocity with magnitude $v$ at an 
angle $\theta$ relative to the Earth's frame will exhibit a proper motion of angular 
displacement $\mu = \tan^{-1}[ \tau  v \sin \theta /\calD]$ in a time period $\tau$. 
Naively, for GLAST, a detection of proper motion 
requires that the angular displacement must be 
greater than the point spread function, i.e., $\mu > \mu_0=9$ arcminutes (which is 
the 68\% PSF for GLAST). However, 
the PSF of GLAST improves to less than 9 arcminutes at energies greater than few GeV because 
the sensitivity of the instrument is not background dominated. It is 
reasonable to assume that a detection of proper motion is defined by the condition 
$\muth > \mu_0 / \sqrt{\Ngth }\approx 4 \,{\rm arcmin}$. 
This is a conservative approach, because 
large number of photon counts (as might be the case for a nearby microhalo) 
may in fact increase the angular resolution of the detector 
\cite{BV99}. 
Furthermore, it is possible that some microhalos are close enough 
that they will have an angular extent (see Fig.~\ref{fig:figure2}) such that the excess 
count in adjacent angular bins may be used to even better localize the source, increasing 
the sensitivity of the detector to proper motion measurements. 
The number of microhalos in a logarithmic interval in $\Mm$ detected with GLAST  
with proper motion greater than $\muth$ is obtained by integrating 
the velocity distribution function with the abundance of microhalos with a velocity 
vector at an angle greater than $\thetamin(\muth,v,\calD) = \sin^{-1}[ \calD \tan \muth / v \tau]$, as
\begin{eqnarray}
\Nm (\Mm, \ge \muth) &=& 4 \pi \int_0^{\dmax(\Mm)} \frac{dN_{\rm m}}{d \calD } \\
 &\times& \left( \int_0^\infty 
f(v) \cos [\theta_{\rm min}(\muth,v,\calD)] d v \right) \, d \calD. \nonumber
\end{eqnarray}

\begin{figure}[t]
\resizebox{!}{7.5cm}{\includegraphics{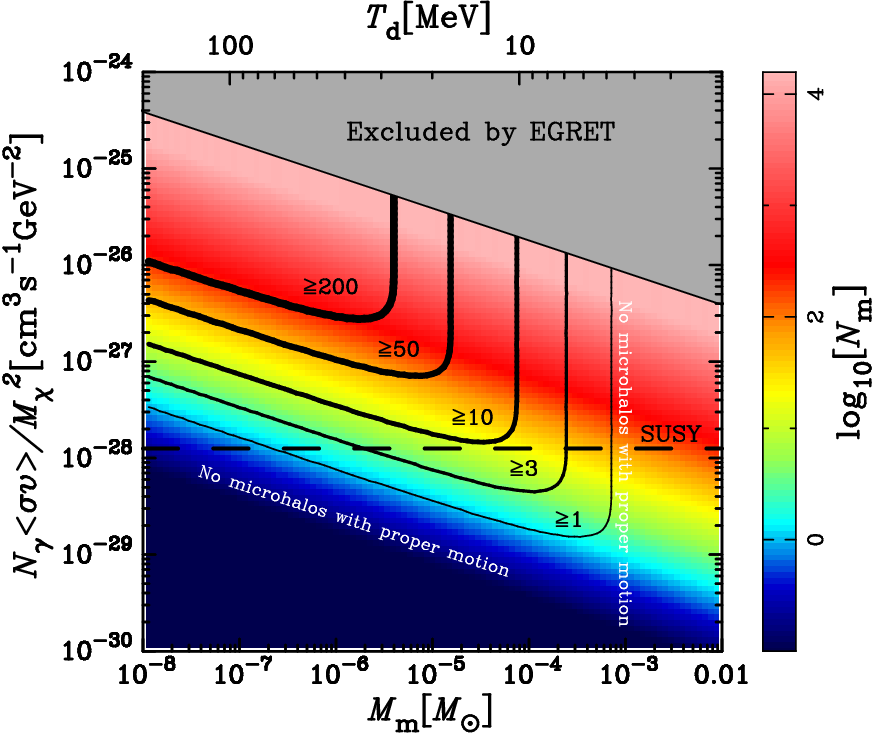}}
\caption {\small {\it Top}: The dependence of the number of visible microhalos per logarithmic mass interval 
to the microhalo 
mass and the properties of the dark matter particle. {\it Solid} lines depict the iso-number 
contours of microhalos which also exhibit detectable proper motion. The {\it dashed} line shows the 
best case scenario for SUSY CDM particle. The shaded area shows the region that is 
already being excluded by EGRET measurements.}
\label{fig:figure1}
\end{figure}

\begin{figure}[t]
\resizebox{!}{7.5cm}{\includegraphics{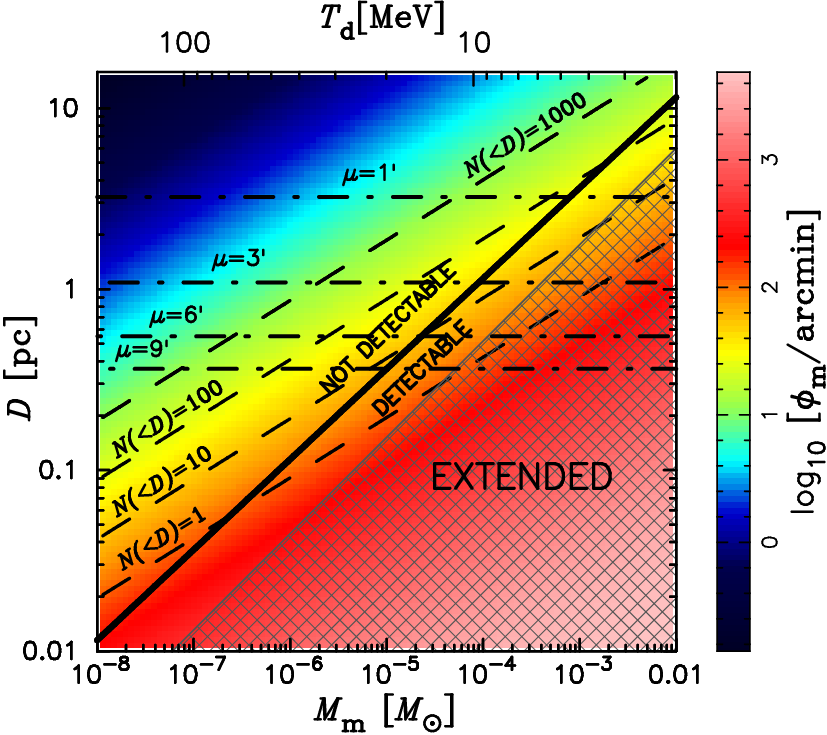}}
\caption {\small The angular size of microhalos as a function of microhalo mass and distance 
for the best case scenario of SUSY CDM. The {\it solid} line defines the distance threshold for 
detection. {\it Dashed} lines depict the 
iso-number contours of the number of microhalos per logarithmic mass interval within a volume defined by $
\calD$. 
{\it Dot-dashed} lines show the maximum distance at which a microhalo could exhibit proper 
motion of 1, 3, 6 \& 9 arcmin. 
The {\it cross-hatched} area corresponds to the 
region where microhalos have an angular extent that should be detectable by 
{\it GLAST}. }

\label{fig:figure2}
\end{figure}

{\it Results}:
Fig.~\ref{fig:figure1} shows the number of visible microhalos per logarithmic mass interval  
and contours of the 
number of microhalos with detectable proper motion for different microhalo masses and 
physics of the dark matter particle. The dashed line depicts the best case scenario 
for SUSY CDM, namely, a neutralino dark matter, with $\Mchi = 46 \GeV$, and 
$\sigmav = 5 \times 10^{-26} \cm^3 \s^{-1}$. The photon spectrum is 
taken from Ref.~\cite{BUB97}, and is integrated from $\Eth = 4 \GeV$. The integration 
time is 2 years.  If, for example, the decoupling temperature of the SUSY CDM 
particle is $\sim 10 \, \MeV$, then microhalos have a mass of $\sim 10^{-4} \, M_\odot$, 
and if $\sim 0.2\%$ of them survive in the Milky Way halo to the present day, 
$\sim 20$ will be detected with GLAST, and 8 out of 20 will exhibit a 
detectable proper motion \footnote{See EPAPS Document No.---- for a similar figure for a 
detector with $\Aeff = 2\times 10^8 \cm^2$ and a PSF of 6 arcmin, and for a figure which 
shows the cumulative number of microhalos with proper motion as a function of the cutoff 
mass in the microhalo mass function in the case of SUSY CDM.  
This document can be reached through a direct link in the online article's HTML 
reference section, or via the EPAPS homepage http://www.aip.org/pubservs/epaps.html}.

\begin{figure}[t]
\resizebox{!}{7.5cm}{\includegraphics{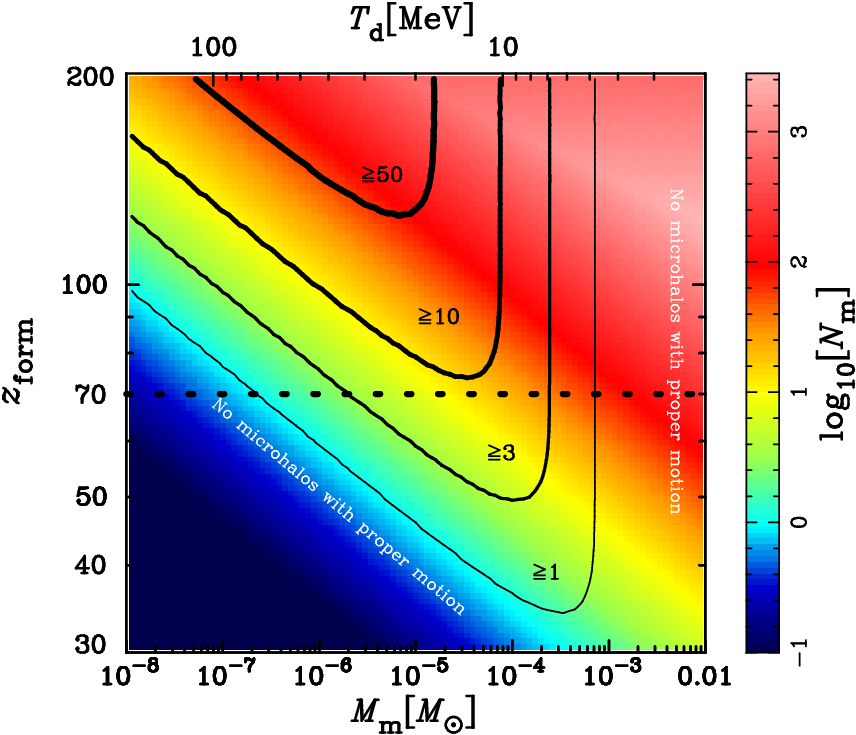}}
\caption {\small The dependence of the number of visible microhalos per logarithmic mass interval 
on the microhalo mass 
and redshift of formation. {\it Solid} lines as in figure~\ref{fig:figure1}. The {\it dashed }
line depicts the redshift at which microhalos form in numerical simulations \cite{DMS05,METAL05,DKM06}.}
\label{fig:figure3}
\end{figure}
The shape of the contours of microhalos with detectable proper motion can 
be understood with the help of Fig.~\ref{fig:figure2}. The radius from the Sun at which a microhalo is 
detected is $\calD \sim \Mm^{1/2}$. However, the radius of the volume that is needed in 
order to encompass at least one microhalo is $\calD \sim \Mm^{1/3}$ (see e.g. the intersection of 
the $N(<\calD)=1$ contour and the detectability line in Fig.~\ref{fig:figure2}). Therefore, for 
small mass microhalos $\Mm \le 10^{-6}\Msun$, the number of microhalos with proper motion 
is limited by the non-abundance of microhalos within a visibility volume. On the other 
hand, for high mass microhalos, even though the detectable number can be high, 
the number of microhalos exhibiting proper 
motion is limited by the distance at which a microhalo can be placed and still exhibit a 
detectable proper motion, because $\mu \sim \calD^{-1}$ (see e.g. the intersection 
of the $\mu = 3'$ and $N(<\calD)=1$ contour in Fig.~\ref{fig:figure2}). 

For smaller(larger) values of $\xi$, such as the case where microhalos are 
disrupted more(less) efficiently, 
the number of visible microhalos and/or ones that exhibit detectable proper 
motion should be multiplied by a factor of $\xi/0.002$. 
An interesting outcome of the detection of a large 
number of microhalos will be the potential measurement of their parallax. 
Such measurements can be used to estimate the 
absolute luminosity and thus 
the measured flux can provide important information about the 
physics of the dark matter particle and its distribution. 
This calculation assumes 
isotropy of the microhalo distribution in the solar neighborhood and thus isotropy in the
distribution of detected microhalos. 
However, the Doppler effect due to the motion of the Earth-Sun system in the 
Milky Way halo will create a dipolar distribution with an amplitude of order 
$\Delta \Nm / \Nm \sim 10^{-3}\%$ \cite{Jordi}. 

In the context of SUSY CDM, 
the detection of at least 1 microhalo with proper motion has an 
important implication: 
the decoupling temperature must be in the range  
$\Td \sim [4-100] \GeV$, and the mass of the particle must be $\Mchi \le 500 \GeV$. 
The former is inferred from the crossing of the SUSY $N_\gamma \langle \sigma v 
\rangle / \Mchi^2$ line with the $\ge 1$ iso-number contour of detectable microhalos 
with proper motion in Fig.~\ref{fig:figure1}. 
The latter is inferred from the lowest value of the quantity $N_\gamma \langle \sigma v 
\rangle / \Mchi^2$ along the same $\ge 1$ iso-number contour in Fig.~\ref{fig:figure1}. 
These bounds can be 
understood in the following way: High decoupling temperatures imply small microhalos, which are less luminous;  
The visibility volume shrinks reducing the number of detectable microhalos 
($\Nm \sim \Mm^{-1}\calD^3 \sim \Mm^{1/2}$). 
On the other hand, low decoupling temperature results in more massive microhalos, 
in a larger visibility volume. Proper motion is inversely proportional to the distance of the microhalo, 
$\mu \sim \calD^{-1} \sim \Mm^{-1/2}$, and so more massive microhalos will tend 
to exhibit smaller proper motions.
It will be of interest to deduce the implications of the kinetic decoupling temperature and particle
mass  
constraints to direct searches for dark matter, because the interactions 
(scalar scattering) that set the kinetic decoupling temperature are the same as the ones that 
are expected to take place in a direct dark matter detection experiment.

The cross-hatched region in Fig.~\ref{fig:figure2} depicts 
the region where microhalos can potentially be resolved 
as extended objects (greater than the 9 arcmin PSF of GLAST). 
If for example microhalos have mass of order $\sim 10^{-3}\Msun$, then there will be $\sim 25$ 
microhalos visible, out of which $\sim 7$ will be resolved as extended objects, and $\sim 2$ 
of them will exhibit a detectable proper motion of greater than $3$ arcmin. This number is a conservative 
estimate since the excess photon flux from adjacent resolution bins in extended objects could be 
used to better localize the position of the 
source\footnote{See EPAPS Document No.---- for examples of the angular profile of photons 
emitted from extended microhalos. 
This document can be reached through a direct link in the online article's HTML 
reference section, or via the EPAPS homepage http://www.aip.org/pubservs/epaps.html}.

Fig.~\ref{fig:figure3} shows the number of visible microhalos and contours of the 
number of microhalos with detectable proper motion for different microhalo masses and 
formation redshift. Recent simulations suggest that microhalos are formed 
at $\zform \sim 70$ (depicted as the dashed line in Fig.~\ref{fig:figure2}). 
The assumption that microhalos form at $\zform \sim 70$ 
should be taken with caution because the redshift of formation depends on 1) the rarity of the 
over-density from which the microhalos collapse \cite{Green:2005fa,DKM06} and 
2) the scale invariance of the primordial power spectrum \cite{Green:2005fa,Bertschinger:2006nq}. 
Furthermore, simulations of small box-sizes are very sensitive to the redshift at which the simulation is 
initiated \cite{Heitmann:2006eu}. This redshift must be high enough such that 
the initial Zel'dovich displacement is a small enough fraction of the mean 
inter-particle separation. For simulations aiming at the formation of 
microhalos this criterion implies a starting redshift much 
higher than what has been assumed in refs.~\cite{DMS05,METAL05,DKM06}. Not fullfiling this 
requirement delays structure formation on all scales in simulations. 
Therefore, if a population of microhalos is formed at a higher redshift than what has been 
assumed in this calculation would imply that 
the number of visible microhalos and microhalos with detectable proper motion 
could be higher as shown in Fig.~\ref{fig:figure2}.

In summary, a detection of the proper motion of $\gamma$-ray sources would be of profound 
importance. 
First and foremost, it will be a detection of annihilating dark matter.
In addition, 
it will provide information on the survival rate of microhalos in the Solar neighborhood 
\cite{DMS05,METAL05,DKM06,ZSH1,Berezinsky:2005py,Green:2006hh}. 
Furthermore, 
it will provide an upper bound to the dark matter particle mass, namely, $\Mchi \le 500$ GeV, an 
area in the parameter space that will be accessible for further studies at the upcoming LHC. 
Moreover, the kinetic decoupling temperature of the dark matter particle must be in the range 
$\Td = [4-100]$ MeV, dramatically restricting the current allowed range of values \cite{PSK06}. 
Therefore, it is essential that a particular analysis for the proper motion of $\gamma$-ray 
sources is performed by the GLAST team, and that the GLAST lifetime is maximized so as to 
increase the baseline for proper motion detection. 

I thank John Beacom for numerous helpful suggestions. 
I acknowledge useful discussions with Kevork Abazajian,
Salman Habib, Katrin Heitmann, Daniel Holz, Gerard Jungman, Louie Strigari,  
Mike Warren and Andrew Zentner. I thank the 
Aspen Center for Physics, where part of this work was completed. 
This work was carried out under the auspices of the NNSA of the U.S. Department of Energy at
Los Alamos National Laboratory under Contract No. DE-AC52-06NA25396.

\bibliography{mu}

\end{document}